\documentclass[showpacs,aps,twocolumn,preprintnumbers,amsmath,amssymb,amsfonts]{revtex4}
\usepackage[dvips]{graphicx}
\newcommand{\si}{\mbox{{\boldmath$\sigma$}}}
\newcommand{\ti}{\mbox{{\boldmath$\tau$}}}

\newcommand{\gr}{\mbox{{\boldmath$\nabla$}}}
\newcommand{\vrad}{\mbox{${\bf r}$}}
\newcommand{\vn}{\mbox{${\bf n}$}}
\newcommand{\vI}{\mbox{${\bf I}$}}
\begin{document}
\begin{titlepage}
\title{$P$ and $T$ odd effects in deuteron in the Reid
potential}
\author{R.V. Korkin}
\email{rvkorkin@mail.ru} \affiliation{Tomsk Polytechnic
University,634034 Tomsk, Russia.}
\begin{abstract}
The $P$ and $T$ odd deuteron multipoles are calculated in the Reid
nucleon-nucleon potential in the chiral limit $m_{\pi}\rightarrow
0$. The contact current generated by the $\pi$-meson exchange does
not contribute to the anapole moment. The contact current
generated by the vector meson exchange is negligible in comparison
with other contributions of vector mesons. The result for the
deuteron electric dipole moment is of great interest because of
the experiment on its measurement discussed in Brookhaven. The
deuteron photodisintegration cross section asymmetry at the
threshold is also calculated. It is shown that its value strongly
depends on the tensor forces and $d$-wave contribution to the
deuteron wave function.
\end{abstract}
\pacs{11.30.Er, 12.15.Mm, 13.40.Ks}
\maketitle
\end{titlepage}
\section{Introduction}
The anapole moment (AM) was introduced by Zel'dovich \cite{zel} as
a very peculiar moment which involved both electromagnetic and
weak interaction. A charged particle interaction with an AM has a
contact nature. Actually, the interaction of electron with nuclei
AM, being of the order of $\alpha G_{F}$, cannot be distinguished
from radiative corrections to the weak interaction due to neutral
currents. The study of Flambaum, Sushkov and Khriplovich \cite{fk}
shows the growth of AM with the nuclei size as $A^{2/3}$. It
means, the parity nonconservation (PNC) interaction of electron
with heavy nuclei AM may become significant and provide important
information about nuclei PNC forces. The competing contribution to
this electron-nucleus interaction -- radiative corrections to the
neutral currents -- does not have an enhancement in heavy nuclei.

The deuteron AM does not have the enhancement as heavy nuclei and
its discussion is possible for another reason -- it has isoscalar
structure only. Radiative corrections to the electron-deuteron
interaction due to $Z$-exchange contains both isoscalar and
isovector contributions. The isoscalar part of interaction which
is operative in deuteron is calculated with sufficiently good
accuracy \cite{ms} and is of the same order of magnitude as the
interaction of electron with the AM. Moreover the deuteron AM is
defined mainly by the $\pi$-meson exchange and is singular in the
$\pi$-meson mass.

The study of PNC effects in deuteron has a long history. On the
one hand, the theoretical predictions are reliable due to small
deuteron binding energy $\varepsilon_{d}\approx 2.23$ MeV. On the
other hand the study of parity violation in the simplest nucleus
-- deuteron -- is very important because of existing discrepancy
between experimental data on PNC forces in $^{133}Cs$ \cite{wi}
and in some other nuclei \cite{ah,des2}.

The deuteron AM was discussed in the series of papers [2,7-11]. In
\cite{kk} the zero range approximation was used to calculate AM.
In \cite{depl1} the deuteron AM was obtained using more realistic
deuteron wave function, based on Argonne $v_{18}$-potential. The
$d$-wave and vector meson contribution in the AM were also
included in the consideration.

The same approach using effective field theory can be applied for
$P$ and $T$ odd moments calculation: electric dipole moment (EDM)
and magnetic quadrupole moment (MQM). The first of them is of a
great interest because of the experiment on its measurement
discussed in Brookhaven. The deuteron EDM and MQM have been
considered in various models in papers [10,12,13].

The last part of the paper is devoted to the $P$ odd asymmetry of
the deuteron photodisintegration cross section calculation in the
Reid potencial \cite{rr}. At present, there is a large dispersion
of the theoretical predictions for this value [15-30]. Most of
them considered various models of weak PNC interaction and
deuteron wave functions. In \cite{korkh} this magnitude was
calculated using the zero range approximation approach and the
obtained value at the threshold was about $A=1\times 10^{-7}$. The
more complicated and reliable approaches used in papers
\cite{depl2} (with Argonne $v_{18}$ potential) and \cite{FT} (with
Paris potential) gave the following answers $A=0.253\times
10^{-7}$ and $A=0.335\times 10^{-7}$ respectively. It seems to be
very interesting to reveal the  nature of the disagreement.

The aim of this paper is to calculate independently $P$ and $T$
odd electromagnetic moments and the deuteron cross section
asymmetry in chiral limit using realistic wave functions obtained
in the soft core Reid potential. It is very important to reveal
the factors which have the greatest influence on $P$ and $T$ odd
moments: $d$-wave contribution and the deuteron wave function
behavior at small distances.

\section{Deuteron wave functions in the Reid potential}

We can obtain the deuteron wave functions using the Reid potential
\cite{rr}. We represent the deuteron $^3S_{1}-^3D_{1}$ state in
the following form:
\begin{equation}
\psi_{d}=\frac{1}{r}\sqrt{\frac{m_{\pi}}{4\pi}}\left(u(x)+\frac{S_{12}}{\sqrt{8}}w(x)\right),
\end{equation}
where $x=m_{\pi} r$ -- the dimensionless distance,
$S_{12}=3(\si_{1}\vn)(\si_{2}\vn)-(\si_{1}\si_{2})$, $m_{\pi}=140$
MeV -- charged $\pi$-meson mass. Normalization condition gives
\[
\int_{0}^{\infty} \left(u^2(x)+d^2(x)\right)d\,x=1.
\]

The Shr\"odinger equation for $u$ and $w$ components can be
written as follows:
\[
u''(x)+ m(E-V_{c}(x))u(x)=2\sqrt{2}m\,V_{t}(x)w(x),
\]
\[
w''(x)+ m(E-V_{c}(x)+2V_{t}(x)+3V_{ls}(x))w(x)
\]
\begin{equation}
=2\sqrt{2}m\,V_{t}(x)u(x).
\end{equation}
Here $V_{c}$, $V_{t}$, and $V_{ls}$ are used for the central,
tensor, and spin-orbit parts of the nucleon-nucleon interaction
potential, respectively:
\[
V(x)=V_{c}(x)+V_{t}(x)S_{12}+V_{ls}(x){\bf L}{\bf S}.
\]

The calculated functions are plotted in Figure 1. The obtained
energy and $d$-wave contribution are: $\varepsilon=2.23$ MeV,
$P_{d}=\int w^2(x)dx=0.065$.

\begin{figure}[h]
\centering \includegraphics[height=5cm]{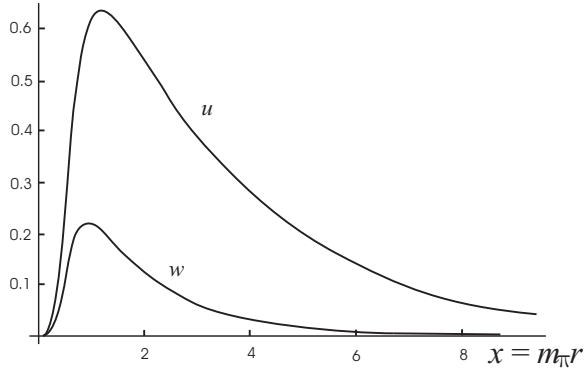}
\caption{Deuteron wave functions $u,w$}
\end{figure}

It should be mentioned that these results can be obtained using
either the soft core or the hard core Reid potential. But in the
future, we will use for calculation of $P$ and $T$ effects the
soft core Reid potential only. The hard core Reid potential is
obviously inappropriate for very small distances description.

\section{AM numerical calculations}

To calculate the perturbed deuteron wave function we will follow
the way based on the direct solution of the Shr\"odinger equation.
Indeed, the perturbed function can be written as
\[
\psi_{d}(\vrad)=\frac{1}{r}\sqrt{\frac{\mu}{4\pi}}\left[u(x)+\frac{S_{12}}{\sqrt{8}}w(x)-i(\si_{1}+\si_{2})\vn
v_{^3P_{1}}(x)\right.
\]
\begin{equation}\label{pert0}
\left.+i(\si_{1}-\si_{2})\vn v_{^1 P_{1}}(x)\frac{}{}\right].
\end{equation}
Functions $v_{^3P_{1}}$ and $v_{^1P_{1}}$ are $p$-waves with the
total spin $1$ and $0$ respectively and the total angular momentum
$1$. To obtain these functions let us use the following weak PNC
nucleon-nucleon potentials due to the $\pi$-meson exchange
\cite{hen}
\begin{equation}\label{pipoten}
V(\vrad)=-i\frac{g\overline{g}}{4\,\pi\,m}\left(\si_{1}+\si_{2}\right)\gr\frac{e^{-m_{\pi}\,r}}{r}
\end{equation}
and $\rho$-, $\omega$-meson exchange \cite{hen}


\[
W=-g_{\rho}\left[h^{0}_{\rho}\ti_{1}\ti_{2}+\frac{1}{2}h^{1}_{\rho}(\tau^{z}_{1}+\tau^{z}_{2})+
\frac{1}{2\sqrt{6}}h^{2}_{\rho}(3\tau^{z}_{1}\tau^{2}_{z}-\ti_{1}\ti_{2})\right]
\]
\[\times\frac{1}{2m}((\si_{1}-\si_{2})\{{\bf
p_{1}-p_{2}},f_{\rho}(r)\}+2(1+\chi_{\rho})[\si_{1}\times\si_{2}]\gr
f_{\rho}(r))
\]
\[
-g_{\omega}[h^{0}_{\omega}+\frac{1}{2}h^{1}_{\omega}(\tau^{z}_{1}+\tau^{z}_{2})]
\]
\[
\times\frac{1}{2m}((\si_{1}-\si_{2})\{{\bf
p_{1}-p_{2}},f_{\omega}(r)\}+2(1+\chi_{\omega})[\si_{1}\times\si_{2}]\gr
f_{\omega}(r))
\]
\begin{equation}\label{poten}
-\frac{1}{2}(\tau^{z}_{1}-\tau^{z}_{2})(\si_{1}+\si_{2})\frac{1}{2m}
\{ {\bf
p_{1}-p_{2}},g_{\omega}h^{1}_{\omega}f_{\omega}(r)-g_{\rho}h^{1}_{\rho}f_{\rho}(r)\}.
\end{equation}

The numerical values of used parameters are listed in Table 1
\cite{ddh}.
\begin{table}[h]
\begin{center}
\begin{tabular}{|c|c|c|c|c|c|c|c|c|} \hline
$g_{\rho}$&$g_{\omega}$&$\chi_{\rho}$&$\chi_{\omega}$&$h^{0}_{\rho}\cdot
10^{7}$&
$h^{1}_{\rho}\cdot 10^{7}$&$h^{2}_{\rho}\cdot 10^{7}$&$h^{0}_{\omega}\cdot 10^{7}$&$h^{1}_{\omega}\cdot 10^{7}$\\
\hline 2.79&8.37&3.7&-0.12&-11.4&-0.2&-9.5&-1.9&-1.1\\ \hline
\end{tabular}
\end{center}
\caption{Numerical values of the constants in potential
\ref{poten}}
\end{table}

Then the Shr\"odinger equation
\begin{equation}
\left(-\frac{1}{m\,r}\frac{d^2}{dr^2}r+\frac{l(l+1)}{m\,r^2}+V\right)\psi_{d}=E\psi_{d}
\end{equation}
can be split into four following equations
\[
u''(x)+ m(E-V_{c}(x))u(x)=2\sqrt{2}m\,V_{t}w(x),
\]
\[
w''(x)+
m(E-V_{c}(x)+2V_{t}(x)+3V_{ls}(x))w(x)=2\sqrt{2}m\,V_{t}u(x),
\]
\[
v_{^3P_{1}}''(x)-\frac{2}{x}v_{^3P_{1}}(x)+m(E-V_{^3P_{1}}(x))v_{^3P_{1}}(x)
\]
\[
=\left[u(x)+\frac{1}{\sqrt{2}}w(x)\right]\frac{\partial}{\partial
x}(F_{\pi}(x)+F^{1}_{\rho}(x)-F^{1}_{\omega}(x))+
\]
\[
+2(F^{1}_{\rho}-F^{1}_{\omega})\left[u'(x)+\frac{1}{\sqrt{2}}w'(x)\right]
\]
\[
-\frac{2}{x}(F^{1}_{\rho}-F^{1}_{\omega})(u(x)-\sqrt{2}w(x)),
\]
\[
v_{^1P_{1}}''(x)-\frac{2}{x}v_{^1P_{1}}(x)+m(E-V_{^1P_{1}}(x))v_{^1P_{1}}(x)
\]
\[
=(u(x)-\sqrt{2}w(x))\frac{\partial}{\partial
x}(3\chi_{\rho}F^{0}_{v}(x)-\chi_{\omega}F^{0}_{\omega}(x))-
\]
\[
-2(3\chi_{\rho}F^{0}_{\rho}(x)-\chi_{\omega}F^{0}_{\omega}(x))\frac{\partial}{\partial
x}(u(x)-\sqrt{2}w(x))
\]
\begin{equation}
+\frac{2}{x}(3\chi_{\rho}F^{0}_{v}(x)-\chi_{\omega}F^{0}_{\omega}(x))(u(x)+2\sqrt{2}w(x))
\end{equation}
with used functions defined as
\[
F_{\pi}(x)=g\overline{g}\frac{e^{-x}}{4\pi x},
\]
\[
F^{0,1}_{\rho}(x)=g_{\rho}h^{0,1}_{\rho}\frac{e^{-\frac{m_{\rho}}{m_{\pi}}
x}}{4\pi x},
\]
\[
F^{0,1}_{\omega}(x)=g^{0,1}_{\omega}\frac{e^{-\frac{m_{\omega}}{m_{\pi}}x
}}{4\pi x}.
\]

The general form of the AM operator is \cite{fk}
\begin{equation}\label{anap}
{\bf a}_{d}=\frac{2\pi}{3}\int d\vrad [\vrad\times[\vrad\times
{\bf j}(\vrad)]]+{\bf a}_{N},
\end{equation}
where ${\bf a}_{N}$ is the nucleon contribution to the deuteron
AM.

Let us consider the AM without nucleon contribution. The current
operator ${\bf j}(x)$ in (\ref{anap}) can be expressed in terms of
the perturbed wave function (\ref{pert0}). Then the AM is
\cite{depl2}
\[
{\bf
a}_{d}=-\frac{2\pi}{3m\,m_{\pi}}\left[\left(\mu_{p}-\mu_{n}-\frac{1}{3}\right)\right.
\]
\[
\times\int_{0}^{\infty}dx\,x(u(x)-\sqrt{2}w(x))v_{^3P_{1}}(x)
\]
\begin{equation}
\left.-(\mu_{p}+\mu_{n})\int_{0}^{\infty}dx\,x\left(u(x)+\frac{1}{\sqrt{2}}w(x)\right)v_{^1P_{1}}(x)\right]e\vI.
\end{equation}

The numerical solution of equations gives the following result for
the AM without nucleon contribution:
\[
{\bf a_{d}}=-\frac{e\vI}{6m m_{\pi}}
\left(14.55\overline{g}+0.048h^{1}_{\rho}-0.132h^{1}_{\omega}\right.
\]
\begin{equation}
\left.-0.074h^{0}_{\rho}-0.051h^{0}_{\omega}\right).
\end{equation}

The nucleon AM (the $pi$-meson exchange only) was calculated in
\cite{kk}
\[
{\bf a_{p}}={\bf
a_{n}}=-\frac{eg\overline{g}}{12mm_{\pi}}\left(1-\frac{6}{\pi}\frac{m_{\pi}}{m}ln\frac{m}{m_{\pi}}\right)\si_{p}
\]
\begin{equation}
=-\frac {e\overline{g}}{12mm_{\pi}} 6.19\,\si_{p}.
\end{equation}

The last numerical result \cite{zhu} was obtained considering
vector meson exchange as well:
\begin{equation}
{\bf
a}_{p,n}=-\frac{e}{12mm_{\pi}}\left(7.61\overline{g}+8.25h^{1}_{\rho}+2.54h^{0}_{\omega}\right)\si_{p,n}.
\end{equation}

We can see that the $\pi$-meson exchange contribution to the
nucleon  AM is very close to our analytical result which we will
use further.

To find the nucleon contribution we should average the sum of the
nucleons AM over the deuteron wave function:
\[
{\bf a}_{N}=\int
\left(u(x)+\frac{S_{12}}{\sqrt{8}}w(x)\right)({\bf a}_{p}+{\bf
a}_{n})
\]
\begin{equation}
\times\left(u(x)+\frac{S_{12}}{\sqrt{8}}w(x)\right)dx=2{\bf
a}_{p}\left(1-\frac{3}{2}P_{d}\right).
\end{equation}

Finally, the total value of the deuteron AM is
\[
{\bf
a}_{d}=-\frac{e}{6mm_{\pi}}(20.14\overline{g}+7.5h^{1}_{\rho}-0.132h^{1}_{\omega}
\]
\begin{equation}\label{anap0}
-0.074h^{0}_{\rho}+1.78h^{0}_{\omega})\vI.
\end{equation}

The contact current due to the $\pi$-meson exchange was obtained
in \cite{kk}
\[
{\bf j}_{c}(\vrad)=\frac{eg\overline{g}}{2\pi
m}\vrad(\vI\gr)\frac{e^{-m_{\pi}r}}{r}.
\]
It was shown also that its contribution to the deuteron AM
vanishes.

The last contribution to be calculated is the vector meson contact
current contribution. In the momentum representation it equals
\[
{\bf j}_{c}=-\frac{\partial W}{\partial {\bf A}}.
\]
But, instead of performing the calculations we can note that this
contribution is suppressed by the factor $\kappa/m_{\rho}\sim
0.06$ in comparison with other vector meson contribution in
(\ref{anap0}). It means, these values are beyond the accuracy of
calculations.

The total constant $C_{2d}$ of $P$-odd electron-deuteron
interaction is the sum of two constants: $C^{r}_{2d}$ -- radiative
corrections to the neutral current and $C^{a}_{2d}$ --
electromagnetic interaction with deuteron AM. The AM constant
$C^{a}_{2d}$ is
\begin{equation}
C^{a}_{2d}=-\alpha
a_{d}\left(\frac{eG_{F}}{\sqrt{2}}\right)^{-1}=0.0075\pm 0.0015.
\end{equation}

We have estimated accuracy on the level about 20\% at the fixed
``best value" $\overline{g}=3.3\times 10^{-7}$ \cite{ddh}.

Combining with the constant due to radiative corrections to the
neutral current \cite{ms}
\[
C^{r}_{2d}=0.014\pm 0.003
\]
we get
\begin{equation}
C_{2d}=0.0215\pm 0.0035.
\end{equation}

The accuracy of this calculation is high enough to give a hope
that the experimental measurement will be able to provide very
useful information about PNC nuclear forces.

\section{Electric dipole and magnetic quadrupole moments}

The calculation of the electric dipole moment (EDM) is of a great
interest because of planned experiments in Brookehaven National
Laboratory.

Let us mention that the smallness of the vector meson exchange
contribution to the deuteron AM is caused by the smallness of the
corresponding coefficients. The deuteron EDM and MQM matrix
elements have very close nature and consequently, vector meson
exchange contribution to them should also be small in comparison
with the $\pi$-meson contribution.

The deuteron EDM calculation can be performed in the same manner
as the AM calculation was done. The perturbed wave function has
the following form:
\[
\psi_{d}(\vrad)=\frac{1}{r}\sqrt{\frac{m_{\pi}}{4\pi}}\left(u(x)+\frac{S_{12}}{\sqrt{8}}w(x)\right.
\]
\begin{equation}
\left.+(\si_{1}+\si_{2})\vn v_{^3P_{1}}(x)+(\si_{1}-\si_{2})\vn
v_{^1P_{1}}(x)\frac{}{}\right).
\end{equation}

The deuteron EDM is
\[
{\bf d}=\langle \psi_{d}| e \vrad_{p}| \psi_{d}\rangle=
\]
\begin{equation}
=\frac{2e}{3m_{\pi}}\int_{0}^{\infty}dx
x\left(u(x)+\frac{1}{\sqrt{2}}w(x)\right)v_{^3P_{1}}(x)\,\vI.
\end{equation}
The simple calculation gives the following value
\[
{\bf d}=-\frac{eg_{1}}{12\pi m_{\pi}}\,5.3\,\vI.
\]
The deuteron magnetic quadrupole moment is
\[
M_{zz}=\frac{e}{3mm_{\pi}}\left[2(\mu_{p}-\mu_{n})\int_{0}^{\infty}dx
x u(x)v_{^3P_{1}}(x)\right.
\]
\[
-2(\mu_{p}+\mu_{n})\int_{0}^{\infty}dx x u(x)v_{^1P_{1}}(x)
\]
\[
-\frac{4\sqrt{2}}{5}\left(\mu_{p}-\mu_{n}-\frac{3}{4}\right)\int_{0}^{\infty}dx
x w(x)v_{^3P_{1}}(x)
\]
\begin{equation}
\left.+\frac{1}{\sqrt{2}}(\mu_{p}+\mu_{n})\int_{0}^{\infty}dx x
w(x)v_{^1P_{1}}(x)\right].
\end{equation}

The numerical calculation gives the following result:
\begin{equation}
M_{zz}=-\frac{e}{12\pi mm_{\pi}}\left(11.5g_{0}+19.8g_{1}\right).
\end{equation}

\section{Cross section asymmetry at the threshold}

Another method to observe $P$ parity nonconservation in deuteron
is measurement of the photodisintegration cross section asymmetry.
Further, we will consider the asymmetry at the threshold of
$\gamma d\rightarrow np$ reaction only because it has the maximum
value there. The experimental result of Leningrad group \cite{lob}
gives for this value
\begin{equation}
A=(1.8\pm 1.8)\times 10^{-7}.
\end{equation}

Unfortunately, this restriction cannot give additional information
about PNC nuclear forces.

It is shown \cite{korkh} that this value is about $10^{-7}$. But
the value calculated is obtained in the zero range approximation
model without $d$-wave mixture consideration. According to [18,19]
these effects are significant and have a large influence on the
final result.

Let us consider the deuteron photodisintegration cross section
asymmetry at the threshold. It is clear that the transition $^3
S_1\rightarrow ^1 S_{0}$ is the only possible at this energy. We
have $M1$ regular transition and $E1$ admixed transition (Fig 2.).

\begin{figure}[h]
\centering \includegraphics[height=4.5cm]{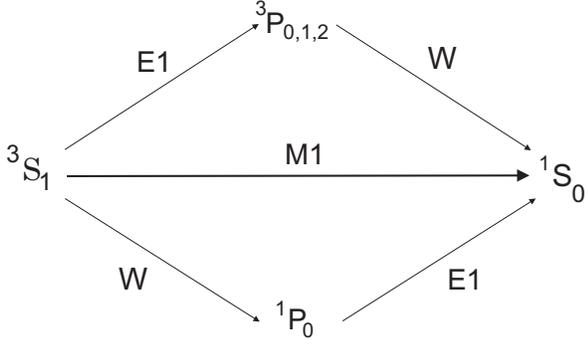} \caption{PNC
transition at the threshold}
\end{figure}

Only the spin nonconservation weak interaction contributes to the
effect. Consequently, the deuteron wave functions and the wave
function of the final $^{1}S_{0}$ state can be written as follows:
\[
\psi_{d}(\vrad)=\frac{1}{r}\sqrt{\frac{\mu}{4\pi}}\left[u(x)+\frac{S_{12}}{\sqrt{8}}w(x)+
i(\si_{1}-\si_{2})\vn v_{^1 P_{1}}(x)\right],
\]
\begin{equation}\label{pert}
\psi_{^1S_{0}}=\frac{1}{r}\left[f(x)+i(\si_{1}-\si_{2})\vn\,g(x)\right].
\end{equation}

To get the value of the asymmetry we should solve the following
system of equations:
\[
u''(x)+
\frac{m}{m_{\pi}^2}(E_{d}-V_{c}(x))u(x)=2\sqrt{2}m\,V_{t}w(x),
\]
\[
w''(x)+
\frac{m}{m_{\pi}^2}(E_{d}-V_{c}+2V_{t}+3V_{ls})w(x)=2\sqrt{2}m\,V_{t}u(x),
\]
\[
v_{^1P_{1}}''(x)-\frac{2}{x}v_{^1P_{1}}(x)+\frac{m}{m_{\pi}^2}(E_{d}-V_{^1P_{1}}(x))v_{^1P_{1}}(x)
\]
\[
=(u(x)-\sqrt{2}w(x))\frac{\partial}{\partial
x}\left(3\chi_{\rho}F^{0}_{v}(x)-\right.
\]
\[
\left.-\chi_{\omega}F^{0}_{\omega}(x)\right)-2(3\chi_{\rho}F^{0}_{\rho}(x)-\chi_{\omega}F^{0}_{\omega}(x))\frac{\partial}{\partial
x}(u(x)-\sqrt{2}w(x))
\]
\[
+\frac{2}{x}(3\chi_{\rho}F^{0}_{v}(x)-\chi_{\omega}F^{0}_{\omega}(x))(u(x)+2\sqrt{2}w(x)),
\]
\[
f''(x)+ \frac{m}{m_{\pi}^2}(E-V_{^1S_{0}}(x))f(x)=0,
\]
\[g''(x)-\frac{2}{x^2}g(x)+\frac{m}{m_{\pi}^2}
(E-V_{^1P_{1}}(x))g(x)
\]
\[
=\frac{\partial}{\partial
x}\left((2+\chi_{\rho})\tilde{F}^{0}_{\rho}(x)+(2+\chi_{\omega})F^{0}_{\omega}(x)\right)f(x)
\]
\begin{equation}\label{diffeq}
+2(\tilde{F}^{0}_{\rho}(x)+F^{0}_{\omega}(x))f'(x)-\frac{2}{x}(\tilde{F}^{0}_{\rho}(x)+F^{0}_{\omega})f(x).
\end{equation}
The following symbols were used:
\[
F_{\pi}(x)=g\overline{g}\frac{e^{-x}}{4\pi x},
\]
\[
F^{0}_{\rho}(x)=g_{\rho}h^{0}_{\rho}\frac{e^{-\frac{m_{\rho}}{m_{\pi}}
x}}{4\pi x},
\]
\[
\tilde{F}^{0}_{\rho}(x)=g_{\rho}
\left(h^{0}_{\rho}-\sqrt{2/3}\,h^{2}_{\rho}\right)
\frac{e^{-\frac{m_{\rho}}{m_{\pi}} x}}{4\pi x},
\]
\[
F^{0}_{\omega}(x)=g_{\omega}h^{0}_{\omega}\frac{e^{-\frac{m_{\omega}}{m_{\pi}}x
}}{4\pi x}.
\]

The asymmetry is
\[
A=-\frac{4m}{3m_{\pi}(\mu_{p}-\mu_{n})\int_{0}^{\infty}f(x)u(x)dx}\times
\]
\[
\times\left(\int_{0}^{\infty}f(x)v_{^1P_{1}}(x)xdx-\right.
\]
\begin{equation}
\left.\int_{0}^{\infty}g(x)(u(x)-\sqrt{2}w(x))xdx\right).
\end{equation}

The numerical value at ``best values'' parameters \cite{ddh} for
$P$-odd nuclear forces is
\begin{equation}
A=0.16\times 10^{-7}.
\end{equation}

This value is much less than the asymmetry calculated in the zero
range approximation with modified wave functions \cite{korkh}. But
the reasonable explanation can be given for this fact. The most
crucial effect on the result has $d$-wave admixture to the
deuteron state. Its influence is much stronger than the wave
functions suppression at small distances. Actually, the same
asymmetry calculated in the Reid potential without $d$-wave
consideration gives the result
\[
A=0.74\times 10^{-7}.
\]

It means that the $d$-wave content in the deuteron wave function
should be determined very precisely to calculate the asymmetry at
the threshold. The Reid potential provides $P_{d}=6.5\%$ for the
$d$-wave content, whereas the same magnitude in Argonne $v_{18}$
potential equals $P_{d}=5.7\%$. This discrepancy can affect the
result very strongly.

\section{Conclusion and comparison}
The deuteron AM calculation based on the phenomenological
nucleon-nucleon interaction was discussed before in papers
[12,13]. The following magnitude was obtained
\[
{\bf a}_{d}={\bf a}_{\pi}+{\bf a}_{nucleon}+{\bf a}_{\rho,\omega}
=-\frac{e}{6mm_{\pi}}\left(14.15\overline{g}+6.96\overline{g}\right.
\]
\begin{equation}
\left.+7.6h^{1}_{\rho}-0.14h^{0}_{\rho}
-0.2h^{1}_{\omega}+2.33h^{0}_{\omega}\right)\vI.
\end{equation}

The $\pi$-meson exchange is splitted here in two terms -- the
contribution of the $\pi$-meson exchange between nucleons and the
additive contribution of the nucleon anapole moments. The
$\pi$-meson contribution is almost the same in our consideration.
The existed discrepancy is beyond the accuracy of both
calculation. As to the vector meson exchange contributions, it
differs from our result up to $50\%$ in some cases. But actually,
this difference has the simple explanation -- the result
dependence on the $d$-wave contribution, which is different for
various potentials. Moreover, as distinct from the $\pi$-meson
exchange, vector meson exchange contribution strongly depends on
the wave function behavior at small distances ($r\le 0.3$ fm),
which makes the results dependent on nucleon-nucleon interaction
model. The Reid potential, as any other potential description as
well, fitting the scattering data up to $350$ MeV cannot give
reliable results for energy $770-780$ MeV needed for the vector
meson exchange description.

Our result for the deuteron EDM is close to that obtained in
\cite{LT}:
\begin{equation}
{\bf d}=-\frac{e}{12\pi
m_{\pi}}(6.06g_{1}+2.37g_{0}+1.05\overline{g}_{\rho}+0.26\overline{g}_{\omega})\vI.
\end{equation}

The main discrepancy concerns the term with $g_{0}$, which
vanishes in our calculations. It could be explained by the fact
that paper \cite{LT} considers terms, non singular in the
$\pi$-meson mass. In our opinion considering this type terms is
beyond accuracy and cannot be justified. The vector meson exchange
contribution to the EDM is small, as it was expected.

The MQM result obtained in \cite{LT} is
\begin{equation}
M_{zz}=-\frac{e}{12\pi mm_{\pi}}(5.62g_{0}+18.6g_{1}).
\end{equation}

We have a reasonable agreement of our result in terms with
$g_{1}$, the discrepancy in another term is large, and cannot be
explained because of lack of details of calculation in \cite{LT}.

The cross section asymmetry in $\gamma d\rightarrow np$ reaction
was calculated in the series of papers. The latest calculations,
based on the realistic deuteron wave functions and the ``best
values'' constants gave the following results
(\cite{depl2},\cite{FT}):
\[
A=0.253\times 10^{-7},
\]
\begin{equation}
A=0.335\times 10^{-7}.
\end{equation}
As we have pointed out this discrepancy can be explained by the
strong $d$-wave influence on the result. Moreover, the above
values show that the calculations of magnitude described by vector
meson exchange as distinct from the $\pi$-meson exchange cannot be
reliably performed.
\begin{center}
***
\end{center}
 I very much appreciate to V.F. Dmitriev and I.B.
Khriplovich for useful and stimulating discussions.


\begin{thebibliography}{99}
\bibitem{zel} Y.B. Zel'dovich, Zh. Eksp. Theor. Fiz. 33, 1531(1957).
\bibitem{fk} V.V. Flambaum and I.B. Khriplovich, Zh. Eksp. Theor. Fiz. 79, 1656(1980).
\bibitem{ms}  W.J. Marciano and A. Sirlin, Phys. Rev. 29 D, 75(1984).
\bibitem{wi}  C.S. Woods et al., Science 275, 1759(1997).
\bibitem{ah}  E. Adelberger and W. Haxton, Ann.Rev.Part. Nucl.Sci. 35, 501(1985).

\bibitem{des2} B. Desplanques, Phys. Rep. 297, 2(1998).
\bibitem{ss}  M.J. Savage and R.P. Springer, Nucl. Phys. A 644, 235(1998), erratum A 657, 457(1999).
\bibitem{ss1}  M.J. Savage and R.P. Springer, Nucl. Phys. A 686, 413(2001).
\bibitem{ss2}  M.J. Savage, Nucl. Phys. A 695, 365(2001).
\bibitem{kk}  I.B. Khriplovich and R.V. Korkin, Nucl.Phys. A 665, 365(2000).

\bibitem{depl1} C.S. Hyun, B. Desplanques, Phys. Lett. B 552, 41(2003).
\bibitem{LT} '.P. Liu and R.G.E. Timmermans, nucl-th/0408060.
\bibitem{posp} O. Lebedev, K.A. Olive, M. Pospelov, and A. Ritz, nucl-th/0402023.
\bibitem{rr}  R.V. Reid, Ann. of Phys. 50, 408(1968).
\bibitem{bs}  R.J. Blin-Stoyle and F. Feshbach, Nucl.Phys. 27, 395(1961).

\bibitem{pa}  F. Partovi, Ann.Phys. (N.Y.) 27, 114(1964).
\bibitem{da1} G.S. Danilov, Phys.Lett. 18, 40(1965).
\bibitem{ta}  D. Tadi\'{c}, Phys.Rev. 174, 1694(1968).
\bibitem{da2} G.S. Danilov, Phys.Lett. B 35, 579(1971).
\bibitem{lee} H.C. Lee, Phys.Rev.Lett. 41, 843(1978).

\bibitem{hh}  W.Y.P. Hwang and E.M. Henley, Ann.Phys.(N.Y.) 129, 47(1980).
\bibitem{hh1} W.Y.P. Hwang, E.M. Henley, and G.A. Miller, Ann.Phys. (N.Y.) 137, 378(1981).
\bibitem{des1} B. Desplanques, Nucl.Phys. A 242, 423(1975).
\bibitem{lmk} K.R. Lassey and B.H.J. McKellar, Phys.Rev.  C 11, 349(1975).
\bibitem{gs}  M. Gari and J. Schlitter, Phys.Lett. B 59, 118(1975).

\bibitem{lmp} J.P. Leroy, J. Micheli, and D. Pignon, Nucl.Phys. A 280, 377(1977).
\bibitem{oka} T. Oka, Phys.Rev. D 27, 523(1983).
\bibitem{korkh} I.B. Khriplovich and R.V.Korkin, Nucl. Phys. A 690, 610(2001).
\bibitem{depl2} C.P. Liu, C.S. Hyun, and B. Desplanques, Phys.Rev. C68, 045501(2003).
\bibitem{FT}   M. Fujiwara, A.I. Titov, Phys.Rev. C69, 065503(2004).

\bibitem{hen} E.M. Henley, Nucl. Phys. A 300, 373(1978).
\bibitem{ddh} B. Desplanques, J.F. Donoghue, and B.R.Holstein, Ann. Phys. (N.Y.), 124,449(1980).
\bibitem{zhu} S.L. Zhu, S.J. Puglia, B.R. Holstein, and M.J. Ramsey-Musolf, Phys. Rev. D 62, 033008(2000).
\bibitem{lob}  V.A. Knyaz'kov, E.A. Kolomensky, V.M. Lobashov, V.A. Nazarenko, A.N.Pirozhkov et al., Nucl. Phys. A 417, 209(1984).
\end{thebibliography}
\end{document}